\documentclass[runningheads]{llncs}
\usepackage[T1]{fontenc}

\usepackage{amsfonts,amsmath}
\usepackage{xcolor}
\usepackage{multirow}
\usepackage{graphicx}

\begin{document}

\title{GA-Unity: A Production-Ready Unity Package for Seamless Integration of Geometric Algebra in Networked Collaborative Applications}
\titlerunning{GA-Unity}

\author{Manos Kamarianakis \and  Nick Lydatakis \and  
George Papagiannakis }
\institute{ FORTH-ICS, University of Crete, ORamaVR S.A.}

\authorrunning{M. Kamarianakis, N. Lydatakis, G. Papagiannakis}

\maketitle

\begin{abstract}
This paper introduces GA-Unity, the first Unity package
specifically designed for seamless integration of Geometric Algebra
(GA) into collaborative networked applications. Indeed, in such
contexts,  it has been demonstrated \cite{kamarianakis2023less} that
using multivectors for interpolation between transmitted poses reduces runtime by 16\% and
bandwidth usage by an average of 50\% compared to traditional
representation forms (vectors and quaternions); we demonstrate that GA-Unity further enhances
runtime performance. Tailored for 3D Conformal Geometric Algebra,
GA-Unity also offers an intuitive interface within the Unity game
engine, simplifying GA integration for researchers and programmers. By
eliminating the need for users to develop GA functionalities from
scratch, GA-Unity expedites GA experimentation and implementation
processes. Its seamless integration enables easy representation of
transformation properties using multivectors, facilitating
deformations and interpolations without necessitating modifications to
the rendering pipeline. Furthermore, its graphical interface
establishes a GA playground for developers within the familiar
confines of a modern game engine. In summary, GA-Unity represents a
significant advancement in GA accessibility and usability,
particularly in collaborative networked environments, empowering
innovation and facilitating widespread adoption across various
research and programming domains while upholding high-performance
standards.
\end{abstract}



\section{Introduction}
\label{sec:introduction}

Geometric Algebra (GA) has garnered significant attention across
various scientific disciplines \cite{hitzer2024current,GAsurvey}, particularly within the realm of
Computer Graphics (CG).  In the domain of CG, Modern 
Game Engines (MGEs) have emerged as
pivotal platforms for application development 
\cite{Jungherr2022The}. 
Among these, Unity stands as a preeminent and widely adopted MGE, 
notably in educational and research contexts \cite{foxman2019united,buyuksalih20173d}.

Higher-dimensional algebras, such as dual quaternions and Geometric
Algebra—specifically, 3D Projective and 3D Conformal Geometric Algebra
(3D PGA and 3D CGA) have demonstrated profound efficacy in CG
applications, particularly in rendering, animation and deformations \cite{hildenbrand2004geometric,dorst2009geometric},
especially in networked environments, as multivector representation of object poses 
leads to more efficient bandwidth usage and better runtime performance \cite{kamarianakis2023less}.

Despite CG experts' proficiency in Euclidean
Geometry and Quaternions, incorporating advanced representations into
applications remains challenging. Key concerns revolve around the
scarcity of readily available, production-ready tools suitable for
integration into MGE environments where applications are developed.
Of particular hindrance is the necessity to construct a Geometric
Algebra (GA) framework from scratch, considering the specialized
functionality required. This includes tasks such as transforming all
points to their GA equivalent multivector form, applying deformations
(e.g., translations, rotations, and dilations), and determining
multivector types. The demand for such GA frameworks is particularly
critical in Computer Graphics, especially in the emerging trend of
collaborative, shared virtual environments
\cite{papagiannakis2008survey,ruan2021networked}. Applications falling
into this category can greatly benefit from the utilization of
multivectors instead of standard representations (matrix and quaternion algebras). Indeed, authors in
\cite{kamarianakis2023less} demonstrated a 16\% runtime improvement
and an average 50\% reduction in bandwidth usage for performing object
interpolations among users collaborating remotely in the same scene.
This highlights the instrumental role that GA can play if adopted by
Unity developers. However, various challenges make this adoption
difficult (see Section~\ref{sub:challenges}).

In response to these challenges, we present GA-Unity. Engineered to
bridge the gap between advanced geometric representations and
practical application development within Unity, GA-Unity offers a
comprehensive solution for CG experts aiming to leverage the power of
Geometric Algebra in their research and development endeavors,
especially in collaborative networked environments. GA-Unity's feature
set includes a pipeline capable of efficiently handling object
transformations in GA forms, both for applying deformations and
interpolating between objects, suitable for real-time visualization.
Significantly, GA-Unity is a production-ready implementation that
addresses performance bottlenecks identified in earlier studies,
ensuring efficiency and scalability in practical networked
applications without compromising compatibility with other networking
pipelines.

The remainder of this paper is structured as follows. In
Section~\ref{sec:related_work}, we provide an overview of previous
work and identify the limitations of existing tools.
Section~\ref{sec:ga_primer} introduces the basic concepts of Geometric
Algebra, laying the foundation for understanding its applications.
Section~\ref{sec:design_architecture} details the design and
architecture of GA-Unity, highlighting its key components. We then delve into the features of GA-Unity in
Section~\ref{sec:ga_unity_features}, discussing how GA-Unity 
can be used for networked collaborative applications, how objects 
are interpolated using GA, and presenting the graphical interface 
of the package.
Section~\ref{sec:performance} evaluates the performance of
GA-Unity, presenting benchmarking results, comparing it with
existing approaches for networked environments. Lastly, in Section~\ref{sec:case_studies}, we
present case studies and applications of GA-Unity in research
environments, game development, and education.

\section{Related Work}\label{sec:related_work}

Previous research has explored diverse approaches to integrating Geometric Algebra (GA) into programming environments, particularly focusing on applications in CG and related fields.

Various Geometric Algebra textbooks, such as those by Hildenbrand et
al. \cite{hildenbrand2004geometric} and Dorst et al.
\cite{dorst2009geometric}, offer introductory insights into Geometric
Algebra, presenting it as a geometrically intuitive framework within
the context of CG. At their core, these texts present
Geometric Algebra as a unified language that facilitates intuitive
object representation and kinematic computation across various
mathematical systems in CG.

In addition to providing a valuable representation, GA implementations
also offer efficiency gains. For instance, Papagiannakis et al.
\cite{papagiannakis2013geometric} demonstrated the efficiency of GA in
real-time character animation blending compared to standard quaternion
geometry implementations. In their study, GA rotors exhibited faster
performance and superior visual quality in real-time character
animation blending scenarios, outperforming traditional quaternion
geometry implementations.

In \cite{Papaefthymiou2016An}, authors provided a CGA-GPU inclusive
skinning algorithm that provides smooth and more efficient results
than standard quaternions, linear algebra matrices, and
dual-quaternions blending and skinning algorithms. Moreover, their
approach avoided conversion between different mathematical
representations, suggesting the implementation of an all-in-one
framework based only in multivectors.

In the same context, Kamarianakis et al. \cite{kamarianakis2021all}
used CGA to perform realistic cuts and tears in rigged character
simulation, enabling new applications in medical surgical simulation.
Their work was further improved in \cite{progressiveTearing}, where
they also used particles to simulate elasticity of the cut/torn model,
for increased realism.

\subsection{Previous Approaches to GA Integrations}\label{sub:previous_approaches}

CLICAL \cite{clical} (1982) was one of the earliest tools for computations involving complex numbers, quaternions, octonions, vectors, and multivectors, enabling geometric, wedge, and dot products.
Gaigen 2 \cite{fontijne2006gaigen} efficiently generates Geometric Algebra (GA) code from high-level algebra specifications, converting them into low-level coordinate-based implementations in various target languages, adapting to program requirements for high performance.
GABLE \cite{mann2001making} and the Clifford multivector toolbox for MATLAB \cite{sangwine2017clifford} offer educational tools for GA in Euclidean 3D-space, supporting Clifford algebras and matrix computations of multivectors.
Garamon \cite{breuils2019garamon}, a C++ library, and Clifford \cite{python_clifford}, a Python module, provide efficient implementations of GA for mixed-grade multivectors, with Garamon employing a prefix tree approach for higher dimensions. The Python module kingdon \cite{kingdon} extends this by supporting multivectors over numpy arrays, PyTorch tensors, or SymPy symbolic expressions with visualization features.
GAALOP \cite{hildenbrand2010gaalop,hildenbrand2020gaalopweb} optimizes geometric algebra files for high-performance parallel computing on platforms like FPGA and CUDA. It integrates with CLUCalc software for interactive GA handling and supports output in formats like C++, OpenCL, CUDA, CLUCalc, or LaTeX.
GAALOPWeb \cite{alves2020efficient}, a Mathematica tool, and another Mathematica package by Aragon et al. \cite{aragon2008clifford}, enhance the manipulation, testing, and visualization of GA algorithms, providing user-friendly interfaces for $n$-dimensional vector space calculations.
Klein \cite{klein}, a C++ library for 3D Projective Geometric Algebra, targets high-throughput applications like animation libraries and kinematic solvers, leveraging SSE for competitive performance without generalizing the space's metric or dimensionality.

\subsection{Challenges in Adopting Existing Solutions in Modern Game Engines}
\label{sub:challenges}

\begin{figure*}
    \centering
    \includegraphics[width=1\linewidth]{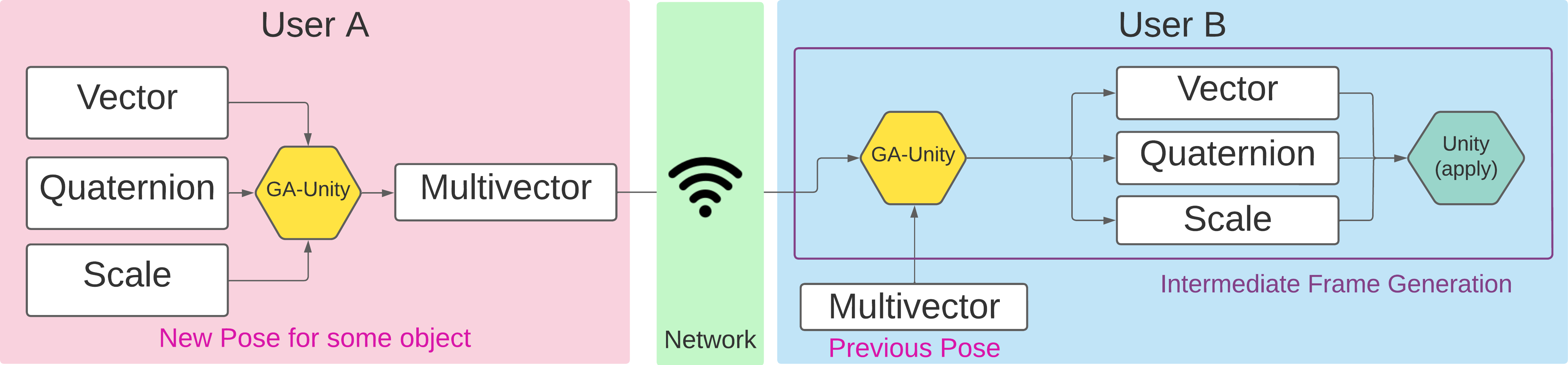}
    \caption{Incorporating GA-Unity within a networked collaborative Unity project. User A sends transformation information to User B over the network in multivector form. GA-Unity is useful for both users: for User A, it streamlines the conversion from typical representation forms (such as vectors, quaternions and scale factors) to GA, while for User B, it aids in decoding and interpolating it. The benefits of this approach is a gain of 50\% in terms of bandwidth \cite{kamarianakis2023less} and 25\% in terms of running performance for the receiving user B, as shown in Section~\ref{sec:performance}.}
    \label{fig:networking}
\end{figure*}

Integrating existing implementations of GA into MGEs  presents several challenges:

\textbf{Language Limitations:} MGEs typically support scripting
languages like C, C\#, or C++, excluding the native execution of
scripts in languages such as Python. This limitation restricts the
availability of readily usable code for application and game
developers, hindering the adoption of GA solutions.

\textbf{Limited Functionality in C\# Packages:} Existing C\# packages,
like those provided generated via
Ganja.js \cite{ganja.js}, offer only basic
multivector functionality. While they provide essential operations
such as geometric, inner, and outer products, conjugation, and scalar
multiplication, they lack comprehensive support for tasks like
defining objects and applying transformations. This forces developers,
including newcomers to GA, to create such functionalities from
scratch, which can be challenging even for experts.

\textbf{Difficulty in Porting Efficient C++ Packages:} Although
packages like Klein, Versor \cite{versor},  written in C++, offer efficiency in terms of
performance, porting them to Unity's C\# environment presents
significant challenges. While it's possible to create a C++-to-C\#
wrapper, this approach introduces complexities such as performance
overhead, maintenance requirements, memory management issues, and
limited access to C++ class interfaces and methods. Moreover, Klein's
limitation to 3D Projective GA precludes the representation of
scalings or round objects like spheres or circles, unlike the more
versatile 3D CGA.

\section{Geometric Algebra Primer}
\label{sec:ga_primer}

\subsection{Fundamentals of Geometric Algebra}
\label{sub:ga_fundamentals}

Geometric algebra is a mathematical framework that unifies and extends
many algebraic systems, including vector algebra, complex numbers, and
quaternions, by introducing the concept of \textit{multivectors}.
Unlike traditional approaches, which treat scalars, vectors, and
higher-dimensional entities separately, geometric algebra provides a
cohesive framework to represent and manipulate these entities
seamlessly. At the heart of geometric algebra is the concept of the
\textit{geometric product}, which generalizes the dot product and the
cross product in Euclidean spaces. Additionally, as all
products can be defined using only the geometric one, we need only use
the latter one along with addition, scalar multiplication, and
conjugation to perform any multivector manipulation.

\subsection{The 3D Conformal Geometric Algebra}
\label{sub:3d_conformal_ga}

In the context of this work we will be employing the so-called
\textit{3D Conformal Geometric Algebra (CGA)}, a 32-dimensional 
extension of dual-quaternions \cite{vinceGeometricAlgebraComputer2008}, 
CGA is a rich mathematical framework that enables the
representation of 3D round elements, such as spheres and circle, as multivectors
within the algebraic structure of CGA. Essentially, CGA serves as the
minimal extension where such representation becomes feasible. 
In the notation of the following sections, we will be using the typical 
basis of CGA, which involves the elements $\{e_i: i=1,2,3,4,5\}$ as well 
as all 32 potential geometric products of up to 5 of them. For convenience, 
we define $e_o$ and $e_\infty$ as $\frac{1}{2}(e_5 -e_4)$ and $e_4+e_5$. The 
inner and outer (or wedge or cross) product of multivectors are denoted by 
$|$ and $\wedge$ respectively. More information 
on the Projective Geometric Algebra (PGA) and CGA can be found in standard GA 
textbooks or notes, such as \cite{dorst2009geometric,Perwass2009,hildenbrand2004geometric}.

\subsection{Geometric Transformations in Conformal Geometric Algebra}\label{sub:deformation}

\begin{table}
    \centering
    \begin{tabular}{|c|c|c|c|}
    \hline
    Type & Translation & Rotation & Dilation\\ 
    \hline\hline
    \multirow{2}{*}{Notes} & \multirow{2}{*}{By $(t_1,t_2,t_3)$} & Equivalent to quaternion & Uniform scale\\
     & &  $q=a-di+cj-bk$ & by $d$ wrt. origin\\
    \hline
    \multirow{2}{*}{Multivector} & $T=$ & $R=$ & $S=$ \\
     & $1-0.5(t_1e_1+t_2e_2+t_3e_3)e_{\infty}$ & $a+be_{12}+ce_{13}+de_{23}$ & $1 + \frac{1-d}{1+d}e_4e_5$ \\
    \hline \rule{0pt}{3ex} 
    Inverse & $T^{-1}=$ & $R^{-1}=$ & $S^{-1}=$ \\
    Multivector & $1+0.5(t_1e_1+t_2e_2+t_3e_3)e_{\infty}$ & $a-be_{12}-ce_{13}-de_{23}$ & $\frac{(1+d)^2}{4d} + \frac{d^2-1}{4d}e_4e_5$ \\
    \hline
    \end{tabular}
    \caption{Multivector forms of translations, rotations and dilations in 3D CGA, as well as their inverses.}
    \label{tab:CGA_forms}
\end{table}

In CGA, translations, rotations and dilations, i.e., uniform scalings with respect to origin, 
can be expressed in multivectors, as shown in Table~\ref{tab:CGA_forms}.

To apply transformations $M_1, M_2, \ldots, M_n$ (in this order) 
to an object $O$, we define the multivector $M := M_n M_{n-1} \cdots M_1$, 
where all intermediate products are geometric. 
The resulting object $O'$ after all transformations are applied is given by:
\begin{equation}
O' = M O M^{-1}
\end{equation}
This represents the final form of $O$ after all transformations have been applied. Notice that 
we standard GA textbooks also use the reverse $\tilde{M}$ instead of the inverse $M^{-1}$
in the equation above, which essentially results in the same object $O'$ potentially multiplied 
by a non-zero scalar; as we are in a projective space, this essentially amounts to the same object.


\section{Design driven by modern needs}\label{sec:design_architecture}

The core objective of GA-Unity is to empower Unity developers,
regardless of their prior experience with Geometric Algebra (GA), to
seamlessly integrate GA into their applications without sacrificing
performance. The design and features of the proposed package revolve
around this central idea.

One of the key features of the GA-Unity package is its native support
for CGA multivectors to represent
geometric relationships between parent and child objects within the
Unity Game Engine. The implementation is crafted to facilitate
developers' rapid comprehension of GA concepts and immediate
visualization of results. Object transformations can be provided
directly in multivector form or generated as multivectors from
mainstream formats such as translation vectors, unit quaternions,
and scale factors. These multivectors can be manipulated and
subsequently utilized by the Unity engine to apply the corresponding
local-to-world transformations to objects.

As previously highlighted in \cite{kamarianakis2023less}, multivectors
are particularly suited for interpolating objects. GA-Unity offers the
capability to interpolate an object between two poses stored in
multivector form (see Section~\ref{sub:interpolation}). The
intermediate transformation multivector obtained through interpolation
is seamlessly applied to the object with increased performance compared to unoptimized implementations
(see Section~\ref{sec:performance}).

Additional features of GA-Unity include the ability to add objects in
multivector form and instantly visualize them (or their duals).
Real-time editing of each coordinate of an object facilitates a deeper
understanding of the geometric implications of coordinate adjustments.
Moreover, GA-Unity supports the creation of more complex objects using
the wedge product for specific object combinations.  
For example, the wedge product of two points with the point at infinity 
$e_\infty$ represents a line passing through these points. Similarly, 
the wedge product of two dual spheres (or planes) corresponds to their 
intersection, which is a circle (or a line, respectively).

\subsection{The GA-Unity Package}\label{sub:components}

The proposed GA-Unity package essentially consists of 4 C\# Unity scripts, described below:

\begin{enumerate}
\item 
\texttt{R410.cs}: Originally generated by Ganja.js, contains
basic multivector class for CGA, along with basic methods such as
scalar multiplication, conjugate, and geometric/inner/outer product.
Also includes basic functions to extract transformation 
information from translators, rotors and dilators, as well as perform linear multivector interpolation.
\item  
\texttt{R410\_Helper.cs}: Used to increase performance by
exposing basis elements outside \texttt{R410}, as a single reference
object. Contains functions that constructs commonly used multivectors
such as translators, rotors and dilations as well as objects such as
points or spheres. 
\item  
\texttt{R410\_pool.cs}: Creates a single reference object that
is used to allocate an extensible pool of multivectors that are used
throughout the interpolation phase. Using multivectors from the pool,
and returning them back once no longer needed, allows avoiding dynamic
allocation of memory and increases performance. 
\item  \texttt{MultivectorLerp.cs}: Contains all the interpolation
pipeline, as described in Section~\ref{sub:interpolation}.
\item  \texttt{GUI.cs}: Contains everything related to the Graphical
User Interface (see Section~\ref{sub:GUI}).
\end{enumerate}

A complete Unity project incorporating GA-enabled deformations and
interpolations only needs this Unity package, along with a scene
containing the objects intended for visualization and/or
interpolation. You can find a minimal open-source working example of
such a project in the GitHub repository (\url{https://github.com/papagiannakis/GA-Unity}), 
along with the necessary documentation. The full closed-source implementation, which improves the
proposed interpolation mechanism for networked collaborative Unity
applications to achieve faster interpolations, has already been
integrated into the MAGES SDK \cite{mages4}, available for free.


\section{GA-Unity Features}\label{sec:ga_unity_features}

\subsection{Simplified Incorporation of GA in Networked Collaborative Applications}

In \cite{kamarianakis2023less}, the authors demonstrated the
significant benefits of using multivector forms to transmit
deformation data in the context of networked collaborative
applications. They showed that employing GA forms leads to a 16\%
reduction in runtime and an average 50\% reduction in bandwidth usage
compared to using mainstream formats such as vectors, quaternions, and
scale factors. This reduction in data transfer across the network
between users is crucial for user immersion, as jittery interpolations
can disrupt it. In collaborative scenarios, such issues can even
compromise application functionality, as objects' positions may not be
updated in time for users to interact with them, leading to situations
like missing hitting a moving ball in a tennis application or failing
to grab an object passed by another user.

While previous works demonstrate the methodology for using
multivectors to alleviate these issues, two major bottlenecks hinder
the adoption of this approach. Developers were required to implement
everything from scratch to incorporate this functionality into their
applications, and an unoptimized implementation could result in performance
hindrances due to continuous memory allocation for 32-dimensional
float arrays representing multivectors. GA-Unity addresses both of
these issues, as it can be directly incorporated into the workflow of
a networked application (see Figure~\ref{fig:networking}).
Furthermore, its design mitigates dynamic yet constant memory
allocation that causes performance overhead using a suitable pooling
mechanism (refer to Sections~\ref{sec:design_architecture} and
\ref{sec:performance}). Finally, the proposed workflow remains 
compatible with existing networking frameworks and pipelines 
such as Photon\footnote{https://www.photonengine.com/}
or Riptide\footnote{https://riptide.tomweiland.net/}.

\subsection{Interpolating Objects in Unity using CGA}\label{sub:interpolation}

Consider the task of interpolating an object between two distinct poses denoted as $P_1$ and $P_2$. These poses are characterized by translation, rotation, and dilation multivectors $(T_i, R_i, D_i)$ for $i=1,2$, respectively. The interpolation factor $\alpha\in[0,1]$ signifies the extent of transition between the poses. At $\alpha=0$, the object aligns with pose $P_1$, gradually transitioning towards pose $P_2$ as $\alpha$ progresses towards 1. Given that objects within MGEs are typically represented as meshes comprising interconnected points, determining the transformed coordinates of each point $x$ within this mesh necessitates consideration of $P_1$, $P_2$, and $\alpha$. This transformed point, expressed in multivector form, is denoted as $x(P_1,P_2;\alpha)$ or simply $x(\alpha)$.

To compute $x(\alpha)$, we apply the requisite transformations $(T(\alpha), R(\alpha), D(\alpha))$ to $x$, facilitating the interpolation between poses $P_1$ and $P_2$ by the factor $\alpha$. Mathematically, this computation is articulated as:
\begin{equation}
x(\alpha) := MxM^{-1} \text{ where } M:=T(\alpha)R(\alpha)D(\alpha).
\end{equation}

\begin{figure*}
    \centering
    \includegraphics[width=1\linewidth]{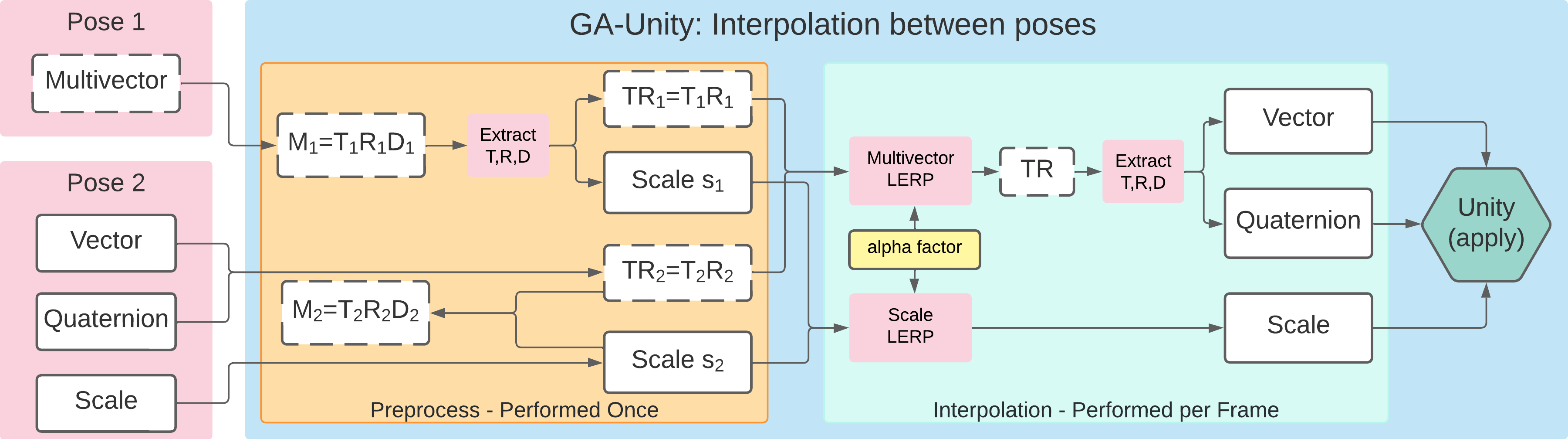}
    \caption{An overview of the proposed pipeline for interpolating between two poses is provided. A pose can be inputted as a single multivector $M$, or as three multivectors $T,R,D$, or via its typical representation form, consisting of a vector, a quaternion, and a scale factor. Following a ``preprocess'' step to extract the $TR$ and scale factor of each pose (if necessary), corresponding vectors, quaternions, and scales can be obtained for each $\alpha$ (alpha factor), which are natively used by Unity. Dashed boxes indicate multivector forms.}
    \label{fig:interpolation_pipeline}
\end{figure*}

The acquisition of these transformations follows a methodology similar
to that elucidated in \cite{kamarianakis2023less}. Specifically, given
the multivectors $M_i=T_iR_iD_i$ for $i\in{1,2}$, we extract 
$D_i$ via the corresponding scale factor $s_i$ using the methodology described in Section~\ref{sub:extractTRD},
and evaluate the multivectors $TR_i:=M_iD_i^{-1} = T_iR_i$. 
We can now perform linear
interpolation with factor $\alpha$ to calculate $TR := (1-\alpha)TR_1 +
\alpha TR_2 = T(\alpha)R(\alpha)$ and again use Section~\ref{sub:extractTRD} to extract $T(\alpha)$ and $R(\alpha)$. $D(\alpha)$ is calculated 
as the dilator corresponding to scale factor $(1-\alpha)s_1+\alpha s)s_2$. Finally, Unity applies the evaluated transformations within its typical pipeline to visualize the object interpolated. A summary of the proposed pipeline is depicted in Figure~\ref{fig:interpolation_pipeline}.

While the resulting interpolation transformations differ from those
obtained using matrix/quaternion representations such as vectors and
quaternions, prior research \cite{kamarianakis2023less} has shown that
the resulting interpolated animations closely match standard outcomes,
especially when the poses are closely situated.

Moreover, one might question why we don't directly evaluate the
interpolated $M:=(1-\alpha)M_1 +\alpha M_2$ and utilize it to derive
$T(\alpha), R(\alpha)$, and $D(\alpha)$. It can be demonstrated that
even a simple interpolation of two dilators does not align with the
linear interpolation of the corresponding scaling factors 
while involving more complex multivectors tends to produce highly non-linear results.

\subsection{Ability to extract transformation from a 
multivector product}\label{sub:extractTRD}

A common mathematical problem that occurs in such applications regards 
the extraction of translation $T$, rotation $R$ and dilation $D$ 
from a \textit{scaling motor}, i.e., a multivector product $M=TRD$.
To solve this problem, we apply $M$ to a unit sphere $C$, centered at origin. The obtained sphere $C' = MCM^{-1}$ is a sphere centered 
at $t$ with radius $d$, that correspond to the translation vector of $T$ and the 
scale factor of $D$ respectively. Since we obtained $T$ and 
$D$, we can then identify $R := T^{-1}MD^{-1}$ and extract the unit 
quaternion using the equivalence in Table~\ref{tab:CGA_forms}. 

Extracting the center and radius of a sphere in multivector
form is quite straightforward, provided we know how it 
is represented in CGA. Indeed, 
a sphere $s$ centered at  $x = (x_1,x_2,x_3)$ with radius $r$ amounts to the CGA multivector
\begin{align}
S =& x_1e_1+x_2e_2+x_3e_3 + \frac{1}{2}(x_1^2+x_2^2+x_3^2-r^2)e_{\infty}+e_o \nonumber\\
=& x_1e_1+x_2e_2+x_3e_3 + \frac{1}{2}(x_1^2+x_2^2+x_3^2-r^2-1)e_4 \nonumber\\
&+\frac{1}{2}(x_1^2+x_2^2+x_3^2-r^2+1)e_5.
\end{align} 
Notice that (a) setting $r=0$ 
would yield the respective multivector for the point $x$ and that (b)
given $S$ we can extract both $x$ and $r$. Indeed, 
if $S[e_i]$ denotes
the coefficient of $e_i$ of a sphere multivector then, 
provided that the sphere is normalized, i.e., $S[e_5]-S[e_4]=1$, we can extract the radius of $r$ of $S$ by evaluating
\begin{equation}
r:= \sqrt{S[e_1]^2+S[e_2]^2+S[e_3]^2-2(S[e_4]+S[e_5])},
\end{equation}
and its center $x$ as it holds that 
\begin{equation}
x:= (x_1,x_2,x_3) = (S[e_1], S[e_2],S[e_3]).    
\end{equation}
If the sphere $S$ is not normalized, we can normalized it
by dividing $S$ with the quantity $S[e_5]-S[e_4]$. The extraction of $x,r$ 
can also be done using GA operations (see eq. 4.75 in \cite{Perwass2009}).

\subsection{A friendly Graphical User Interface for Multivector Manipulation}
\label{sub:GUI}

To further enhance GA adoption among Unity developers, we have
developed a simple yet powerful Graphical User Interface (GUI) inspired by tools
like Geogebra \cite{geogebra}. This GUI,
streamlines the process of
adding multivector objects such as lines, circles, and planes, while
also enabling the creation of new objects through combinations (e.g.,
intersections or unions) of existing ones. Users can effortlessly add
an object (along with its \textit{dual}
\cite{hildenbrand2004geometric}), which is stored in multivector form
and visualized using Unity. The object's coordinates can be edited in
real-time, providing an additional educational layer as users gain a
better understanding of the geometric properties associated with each
coefficient.

Creating objects via intersections, joins, or meets of other objects
allows users to grasp the geometric power inherent in CGA. For
instance, users can define a line in CGA by joining two points or a
circle by joining three points. Additionally, users have the option to
deform objects or interpolate them between two poses, utilizing the
methodologies described in Sections~\ref{sub:deformation} and
\ref{sub:interpolation}.

Regardless of their actions, the related multivectors are always
displayed and stored, ready to be reused. We anticipate that this
comprehensive approach will greatly enhance the accessibility of GA,
providing an educational playground for GA within the familiar
environment of an MGE like Unity.

\section{Performance Evaluation in Networked Environments}\label{sec:performance}

\begin{table}
    \centering
    \begin{tabular}{|c||c|c||c||c|}
        \hline
        Objects  & Previous  & Our  & Improvement & Improvement  \\
        Used   & Method\cite{kamarianakis2023less} & Method & Ours VS \cite{kamarianakis2023less} & Ours VS vectors \& quaternions \\
        \hline
        50   & 1,33 ms  &  0,97 ms 	  &	27\% & 20\% \\
        100  & 2,65 ms  &  1,00 ms 	  &	62\% & 26\% \\
        150  & 3,90 ms  &  1,13 ms 	  &	71\% & 27\% \\
        200  & 5,27 ms  &  1,19 ms 	  &	77\% & 28\% \\
        250  & 6,44 ms  &  2,62 ms 	  &	59\% & 25\% \\
        \hline
    \end{tabular}

    \caption{(Columns 1-4) Performance comparison between our GA implementation and the previous GA implementation \cite{kamarianakis2023less}. The metrics indicate the time required to perform the interpolation of a set of objects, each with varying cardinality. Our implementation demonstrates a performance boost of over 20\% as the number of interpolated objects increases. Results were 
    obtained using only CPU operations in a Windows 10, Intel Core i5-8500 3.00 GHz machine. (Column 5) Comparison performance of our method compared to typical pipeline using vectors for transformation and quaternions for rotations. 
    The 16\% increased performance demonstrated between the typical pipeline and the method proposed in \cite{kamarianakis2023less} is further enhanced by the percentages in column 4. Notice that the performance boost
    was obtained by sending less information per objects per second in multivector form.}
    \label{tab:performance_gain_compared_with_less}
\end{table}

The previously proposed implementation for networked applications
\cite{kamarianakis2023less} required the creation of new multivectors
at each interpolation state, leading to frequent memory allocations
for 32-dimensional vectors, often occurring multiple times per second
to generate frames.  As transformation multivectors only use at most 
16 floats, 16-dimensional arrays were used instead. However, even with 
this optimization, the memory allocation
process posed a performance bottleneck, which we addressed by
implementing a pooling mechanism approach.
Table~\ref{tab:performance_gain_compared_with_less} illustrates that
our approach resulted in an average decrease of more than 50\% in the
time required to perform the necessary interpolations for a set of
objects with varying cardinality. It is evident that as more objects
are interpolated, the performance gain becomes more pronounced, up to
a certain point. This gain occurs for each user that receives and
interpolates a multivector, as illustrated by User B in
Figure~\ref{fig:networking}. Notably, this gain is on top of the
previously reported 16\% benefit that occurred simply by replacing
traditional with GA-based forms \cite{kamarianakis2023less} and sending 
less transformation information per frame to achieve the same visual result. In
conclusion, in networking collaborative applications, GA-Unity may
provide an average improvement of more than 25\% compared to
traditional approaches, regarding runtime.  

To assess the performance of the interpolation, we conducted the following steps. 
Initially, we determined the scene's initial frame rate. For example,
if we measured 250 FPS, it implies that we required 4 (1000/250)
milliseconds (ms) to execute the necessary operations for each
frame. Subsequently, we evaluated the frame rate of the same scene
while continuously interpolating a given number of cubes (50,100, 150, 200 or 250) over an
average duration of 10 seconds.
Again, we identified the required time to evaluate its frame using 
the recorded FPS. Finally, we computed the difference in time required for
each frame, which represents the time needed to calculate the objects'
interpolation. 

As a final remark, the preliminary performance analysis presented aims
to demonstrate the improvements over previous implementations, rather
than offering a comprehensive benchmark of the tool, which falls
outside the scope of this paper.


\section{Case Studies and Applications}\label{sec:case_studies}
\subsection{Use Cases in Research Environments}\label{sub:research_use_cases}

MGEs such as Unity are commonly utilized in research settings to
emulate complex environments, apply diverse physics laws, and simulate
scenarios based on user-defined parameters. This widespread adoption
stems from the need for realistic visualization of experiments and
work, coupled with the ability to efficiently solve or approximate
complex equation systems. 

GA-Unity seamlessly integrates into such projects without apparent
rendering performance. Moreover, it can complement various physics
calculations, leveraging the effectiveness of GA in diverse scientific
domains \cite{GAsurvey}. The ease of incorporating GA into Unity-based
projects could further accelerate its adoption, making it more
prevalent within the research community.

\subsection{Using GA for game development \& industrial large-scale projects}\label{sub:real_time_visualization}

Visualization tools such as Ganja.js and CLUCalc offer representations
of deformations or objects using Geometric Algebra (GA). However,
these options lack the capability to visualize complex scenes
realistically. They fall short in providing features like shadows,
lighting options, or automatic object animation. Additionally, they
struggle to scale up and integrate seamlessly into MGEs, limiting their usefulness in game development and/or 
large-scale projects tailored to industry demands.

GA-Unity addresses these limitations by enabling the integration of GA
into production-ready applications and large-scale products,
leveraging the extensive usage of Unity in their development.
This package facilitates the swift adoption of multivector representation
forms, thereby introducing GA to a rapidly expanding market, akin to
the widespread adoption of quaternions in the past.

One of the key advantages of GA-Unity is its seamless integration with
Unity's native support for animations. It replaces typically used
transformation representation forms with GA equivalents, allowing for efficient
interpolations. This transition is particularly beneficial in
collaborative networked Virtual Reality (serious) games and 
applications, where constant
exchange and interpolation of transformations among users are
essential \cite{kamarianakis2023less}.

\subsection{Educational Implementations}\label{sub:educational_implementations}

Unity is widely recognized for its use in undergraduate and graduate
CG curricula worldwide. Typically, computer scientists
in these courses are introduced to fundamental representation forms
such as transformation matrices and quaternions. However, apart from a
few exceptions \cite{papagiannakis2014glga,elements}, they seldom
delve deeper into advanced geometric forms like geometric algebra
and/or dual-quaternions. One major reason for this gap is that the
Unity platform used for development lacks native support for such
forms, making it challenging, if not impossible, for students to
understand and integrate GA seamlessly. GA-Unity addresses this
limitation by enabling the early adoption of GA in CG
educational curricula, thereby significantly impacting the
proliferation of GA knowledge. 

For example, using GA-Unity in a undergraduate CG course at University 
of Crete, students were able to easily perform a simple task 
of cube interpolation between two poses using GA, where they 
better understood the power of using alternative representation 
forms.


\section{Conclusions, Future Work and Acknowledgements}\label{sec:conclusions_future_work}

The incorporation of Geometric Algebra (GA) into the Unity Game Engine
via the open-source GA-Unity package, available at https://github.com/papagianna\-kis/GA-Unity, marks a significant milestone in the realm of
CG and simulation.In the context of collaborative networked
applications, utilizing GA for representing, exchanging, and
interpolating transformation data offers advantages in both bandwidth
utilization and runtime performance. In this study, we've showcased
that runtime performance can be further enhanced from the previously
reported 16\% \cite{kamarianakis2023less} to over 25\%, while the
bandwidth benefits remain consistent at an average of
50\%. Furthermore, by designing GA-Unity to seamlessly integrate with
the Unity pipeline and releasing it as an open-source tool, we aim to
democratize GA for developers and help them explore the benefits of GA
within the familiar environment of the Unity game engine.

In our future work, we aim to enhance performance by
introducing GPU or parallel computations as well as incorporate 
production-ready C++ framework such as Klein. We will investigate the benefits
of using various GA's such as 3D PGA or $G_{6,3}$ as well as alternative 
interpolation techniques such as logarithmic multivector blending. 
Lastly, we envision adapting GA-Unity to various MGEs beyond Unity, such as 
Godot\footnote{https://godotengine.org/} or even custom-made game engines.

This work is partially supported by the OMEN-E project (PFSA22-240), that have received 
funding from the Innosuisse Accelerator programme.

\bibliographystyle{splncs04}
\bibliography{references}

\end{document}